\documentclass[sigconf]{acmart}
\graphicspath{{figure/}}
\usepackage{multirow}
\usepackage{colortbl}
\definecolor{RGCDblue}{RGB}{232,235,255}
\AtBeginDocument{%
  }

\setcopyright{none}
\copyrightyear{2018}
\acmYear{2018}
\acmDOI{XXXXXXX.XXXXXXX}
\acmConference[Conference acronym 'XX]{Make sure to enter the correct
  conference title from your rights confirmation email}{June 03--05,
  2018}{Woodstock, NY}
\acmISBN{978-1-4503-XXXX-X/2018/06}
\renewcommand\footnotetextcopyrightpermission[1]{}

\begin{document}

\title[Bridging Short Videos and Live Streams]{Bridging Short Videos and Live Streams: Reasoning-Guided Multimodal LLMs for Cross-Domain Representation Learning}

\author{%
Le Zhang\textsuperscript{1,*}, Xiaolan Zhu\textsuperscript{1,*}, Yuchen Wang\textsuperscript{1,*}, Shilong Kang\textsuperscript{1}, Jiaqi Xue\textsuperscript{1},\\
Xiaoyu Zhang\textsuperscript{1}, Xiang Chen\textsuperscript{1}, Yalong Guan\textsuperscript{1}, Xiangyu Wu\textsuperscript{1},\\
Shijun Wang\textsuperscript{1,\textdagger}, Lantao Hu\textsuperscript{1}, Kun Gai\textsuperscript{1}\\[-1pt]
\normalsize \textsuperscript{1}Kuaishou Technology, Beijing, China\\[-1pt]
\normalsize \{zhangle07, zhuxiaolan, wangyuchen11, kangshilong, xuejiaqi03,\\[-1pt]
\normalsize zhangxiaoyu, chenxiang08, guanyalong, wuxiangyu06, wangshijun03,\\[-1pt]
\normalsize hulantao\}@kuaishou.com, kun.gai@qq.com}
\makeatletter
\g@addto@macro\@authornotes{%
  \stepcounter{footnote}\footnotetext{Equal contribution.}%
  \stepcounter{footnote}\footnotetext{Corresponding authors.}}
\makeatother

\renewcommand{\shortauthors}{Zhang et al.}

\makeatletter
\def\@mkbibcitation{\bgroup
  \let\@vspace\@vspace@orig
  \let\@vspacer\@vspacer@orig
  \def\@pages@word{\ifnum\getrefnumber{TotPages}=1\relax page\else pages\fi}%
  \def\@article@string{\ifx\@acmArticle\@empty{\ }\else,
    Article~\@acmArticle\ \fi}%
  \par\medskip\small\noindent{\bfseries ACM Reference Format:}\par\nobreak
  \noindent Le Zhang, Xiaolan Zhu, Yuchen Wang, Shilong Kang, Jiaqi Xue,
  Xiaoyu Zhang, Xiang Chen, Yalong Guan, Xiangyu Wu, Shijun Wang,
  Lantao Hu, and Kun Gai. \@acmYear. \@title
  \ifx\@subtitle\@empty. \else: \@subtitle. \fi
  \if@ACM@nonacm\else
    \if@ACM@journal@bibstrip
       \textit{\@journalNameShort}
       \@acmVolume, \@acmNumber \@article@string (\@acmPubDate),
       \ref{TotPages}~\@pages@word.
    \else
       In \textit{\@acmBooktitle}%
       \ifx\@acmEditors\@empty\textit{.}\else
         \andify\@acmEditors\textit{, }\@acmEditors~\@editorsAbbrev.%
       \fi\
       ACM, New York, NY, USA%
         \@article@string\unskip, \ref{TotPages}~\@pages@word.
    \fi
  \fi
  \ifx\@acmDOI\@empty\else\@formatdoi{\@acmDOI}\fi
\par\egroup}
\makeatother

\begin{abstract}
As live streaming services grow, many platforms offer short videos and live streams to meet diverse needs. Short videos carry substantial traffic and rich behavior signals, whereas live streaming is a core conversion scenario with sparse behavior data, making cold start severe. Transferring user interests from short videos to live streaming recommendation can alleviate these issues. Meanwhile, short videos and live streams are complex multimodal items, and integrating multimodal signals improves recommendation performance. Although Multimodal Large Language Models (MLLMs) show strong multimodal understanding and reasoning, their application to cross-domain recommendation remains underexplored.
To this end, we propose Reasoning-Guided Cross-Domain Representation Learning (\textbf{RGCD-Rep}), a reasoning-guided framework for cross-domain recommendation from short videos to live streams. RGCD-Rep introduces MLLM reasoning resource-efficiently and learns transferable item representations guided by behavioral collaboration via two-stage training. First, reasoning-aware distillation lets a frozen teacher MLLM generate structured cross-domain reasoning knowledge and distills it into a lightweight student MLLM. Second, transferability-guided cross-domain representation learning decomposes item representations into transferable and domain residual representations. The resulting representations are computed offline and integrated into downstream retrieval tasks, enabling low-cost industrial deployment. Extensive offline experiments demonstrate RGCD-Rep's superiority. After deployment in Kuaishou's live streaming recommendation system, A/B tests show significant gains across multiple core business metrics, confirming its effectiveness and practicality in real industrial scenarios. RGCD-Rep is fully deployed and serves over $400$ million users daily.
\end{abstract}



\maketitle

\section{Introduction}
With mobile internet technologies advancing rapidly, live streaming has grown in
recent years \cite{guo2026room, qu2025bridging, cao2026foresight}. Meanwhile,
more platforms, such as YouTube, TikTok, and Kuaishou, provide short videos
alongside live streams to meet users' and content creators' diverse needs. In
platform ecosystems, short videos carry substantial traffic with active user
behaviors and rich interaction data, whereas live streaming serves as a core
conversion scenario where users interact in real time through virtual gifts and
other actions, contributing significantly to platform revenue. However, due to
exposure differences across content domains, live streaming behavior data is
usually sparse, and high-value behaviors such as follows and virtual gifting are
particularly scarce \cite{LiYangWenEtAl2025FARM, deng2024mmbee}. In addition,
cold start is especially severe for new streamers and users.

To alleviate data sparsity and cold-start issues in live streaming, a
straightforward solution is cross-domain recommendation(CDR), which connects
short videos with live streams and transfers user interests from short
videos to live streaming recommendation \cite{zhao2023crossdomain,
li2024aiming, li2025disco}. Existing studies mainly improve CDR from three
perspectives. First, transfer modules integrate source-domain representations
into target-domain recommendation
\cite{HuZY18CoNet,ManShenJinCheng2017CrossDomain}. Second, graph neural
networks \cite{LiuLLP20BiTGCF,XieLiuWangLiuZhangLin2022CCDR} and sequential
models capture cross-domain behavioral dependencies
\cite{CaoCongShengLiuWang2022C2DSR,HuXiaZhangFuWuHuanLiTangZhou2024EnhancingSR}.
Third, with increasingly available visual, audio, and textual signals,
item-level multimodal content is incorporated to enhance CDR \cite{DaiEtAl2024UniEmbedding,YangYL2023MOTKD}. For example,
MOTKD \cite{YangYL2023MOTKD} introduces a multimodal graph attention network to
model user preferences and transfer them to the target domain.
UniEmbedding \cite{DaiEtAl2024UniEmbedding} extracts item representations via a
domain-aware multimodal adapter and a user-view projection module.

Since live streams and short videos are complex items with visual, audio, and
textual modalities, incorporating multimodal content is crucial for CDR.
However, existing studies commonly rely on pre-extracted modality-specific
representations for item modeling \cite{DaiEtAl2024UniEmbedding,YangYL2023MOTKD}.
\textbf{Such static features are limited in deep semantic reasoning over multimodal content 
and provide weak explanations for why items from two domains can be behaviorally transferable.}
These limitations motivate exploring MLLMs for
multimodal CDR. MLLMs integrate multimodal information into a unified textual
semantic space, improving recommendation systems' ability to understand and
explain complex multimodal inputs \cite{liu2023llava, dai2023instructblip}.
This offers a promising way to model complex cross-domain user behaviors and
improve recommendation accuracy.
However, MLLM-enhanced CDR still faces three major challenges.
\textbf{Ch1:} Large-scale MLLMs incur high computational costs, making direct industrial
deployment impractical \cite{caffagni2024revolution, ye2025harnessing}.
\textbf{Ch2:} Relying only on MLLMs' open-world knowledge may yield
task-irrelevant or inaccurate interpretations \cite{ye2025harnessing}.
\textbf{Ch3:} User behavior patterns in CDR are complex, and cross-domain
interests are not always consistent. Directly aggregating shared
and domain-specific information from both domains may thus cause negative
transfer \cite{zhu2022ptupcdr,park2023negative}.

To address these challenges, we propose RGCD-Rep, a reasoning-guided
cross-domain representation learning method. RGCD-Rep introduces MLLM reasoning
resource-efficiently and learns generally transferable item
representations.
\textbf{To address Ch1}, we design a reasoning-aware distillation strategy that transfers the reasoning capability of a large teacher MLLM to a lightweight student MLLM.
Specifically, CoT prompting guides the teacher to generate structured reasoning
knowledge, while distillation enables the student MLLM to learn step-by-step
reasoning patterns and perform structured cross-domain reasoning.
\textbf{To address Ch2}, we design a behavior-grounded two-stage
training framework. By constructing behavior-grounded cross-domain samples and learning
objectives, the lightweight model is fine-tuned
end-to-end. Thus, the student model inherits teacher reasoning ability and
captures cross-domain behavioral co-occurrence signals.
\textbf{To address Ch1 and Ch3}, we further design a
transferable-residual query-aware aggregation module on top of the student MLLM,
which decomposes item representations into transferable and residual components. The resulting transferable representations can be
generated offline and directly integrated into downstream retrieval tasks,
enabling low-cost industrial deployment.
We summarize our main contributions as follows:

\begin{list}{\labelitemi}{%
    \setlength{\leftmargin}{1.5em}%
    \setlength{\labelwidth}{1em}%
    \setlength{\labelsep}{0.5em}%
    \setlength{\itemsep}{0pt}%
    \setlength{\parsep}{0pt}%
}
    \item To the best of our knowledge, this is the first study to introduce
    MLLM reasoning into CDR, offering a new perspective on transfer among
    complex multimodal items.
    \item We propose RGCD-Rep, a reasoning-guided cross-domain representation
    learning framework. RGCD-Rep transfers CoT reasoning from a large teacher
    MLLM to a lightweight student MLLM via reasoning-aware distillation and
    learns behavior-grounded, generally transferable item representations through
    end-to-end fine-tuning, enabling resource-efficient industrial deployment.
    \item Extensive experiments on industrial datasets and online deployment in
    a real recommendation system demonstrate the effectiveness and practicality
    of RGCD-Rep, with significant gains in both offline performance and online
    business metrics.
\end{list}

\section{Related Work}
\textbf{Cross-domain Recommendation.} CDR aims to transfer useful knowledge from a source domain to a target domain to improve target-domain recommendation performance,
effectively alleviating cold-start and data sparsity issues.
Early studies mainly transfer collaborative signals by sharing user representations or modeling cross-domain user--item interactions.
For example, CoNet~\cite{HuZY18CoNet} introduces cross connections between the networks of two domains to enable cross-domain knowledge transfer.
PTUPCDR~\cite{zhu2022ptupcdr} learns personalized preference transfer functions that adaptively transfer user preferences from the source domain to the target domain.
Subsequent studies employ graph neural networks and sequential models to capture user behaviors across multiple domains.
C2DSR~\cite{CaoCongShengLiuWang2022C2DSR} constructs a cross-domain item transition graph based on item co-occurrence relations,
thereby capturing associations between items from different domains.
From the perspective of user sequence modeling,
Tri-CDR~\cite{ma2024tricdr} designs multiple contrastive learning tasks to capture fine-grained global interests from mixed behavior sequences.
In live streaming recommendation,
recent studies begin to exploit short video behaviors to alleviate the sparsity of live streaming interactions.
FARM~\cite{LiYangWenEtAl2025FARM} transfers user preferences from the short video domain to the live streaming domain,
and MGCCDR~\cite{qu2025bridging} further constructs multi-graph contrastive learning signals between short videos and streamers,
demonstrating the effectiveness of short video behaviors for live streaming recommendation.
Although these methods bridge different domains through behavior sequences,
item co-occurrence,
or graph structures,
they are still mainly driven by collaborative signals and provide limited semantic explanations for why cross-domain items are transferable.

\textbf{Multimodal Recommendation.} With the rapid growth of visual,
textual,
and audio content on online platforms,
multimodal recommendation has been widely studied for user preference modeling and item representation learning.
Early studies mainly incorporate multimodal features as auxiliary information into recommendation models.
For example,
VECF~\cite{chen2019vecf} employs a visual encoder to capture image-region information for user preference modeling.
while ACF~\cite{chen2017acf} designs item- and component-level attention over visual features for multimedia recommendation.
With the development of graph neural networks,
many studies further exploit high-order user--item relations for multimodal preference learning.
MMGCN~\cite{wei2019mmgcn} and GRCN~\cite{wei2020grcn} construct modality-aware user--item graphs or multimodal interaction graphs to model preferences from different modalities.
Recently,
multimodal information has also been introduced into CDR.
MOTKD~\cite{YangYL2023MOTKD} introduces a multimodal graph attention network to model users' multimodal preference representations and transfer them to the target domain.
UniEmbedding~\cite{DaiEtAl2024UniEmbedding} learns general multimodal multi-domain item representations through a domain-aware multimodal adapter and a user-view projection module.
These studies show the effectiveness of multimodal information in recommendation and CDR.
However,
most existing methods directly adopt pre-extracted multimodal features,
or learn multimodal representations through shallow alignment and feature fusion.
They lack explicit semantic reasoning over multimodal content and provide limited explanations for why items from different domains can be behaviorally transferable.

\section{Preliminaries}

We focus on multimodal CDR between the short video domain and the live streaming domain,
where the short video domain serves as the source domain and the live streaming domain serves as the target domain.
In principle, the proposed method is also applicable to knowledge transfer across multiple domains.
Let $\mathcal{U}=\{u_1,u_2,\ldots,u_{N_U}\}$ denote the user set,
$\mathcal{L}=\{l_1,l_2,\ldots,l_{N_L}\}$ denote the live streams and
$\mathcal{S}=\{s_1,s_2,\ldots,s_{N_S}\}$ denote the short videos,
where $N_U$, $N_L$, and $N_S$ are the numbers of users, live streams, and short videos, respectively.
We define $\mathbf{E}_{UL}\in\{0,1\}^{N_U\times N_L}$ with entries
$\{e_{ul}\mid u\in\mathcal{U}, l\in\mathcal{L}\}$ as the user interactions in the live streaming domain,
and $\mathbf{G}\in\{0,1\}^{N_U\times N_S}$ with entries
$\{g_{us}\mid u\in\mathcal{U}, s\in\mathcal{S}\}$ as the user interactions in the short video domain.
Here, $e_{ul}, g_{us}\in\{0,1\}$, and a value of $1$ indicates that the corresponding user and item interact with each other.
In addition, each item contains multimodal content, including video frames, OCR/ASR transcripts, title text, and other metadata.
The goal of this paper is to learn a general cross-domain transferable representation that captures user interest factors
shared between the short video and live streaming domains while filtering domain residual information that may lead to negative transfer.

\section{Methodology}
\subsection{Overview}
\begin{figure*}[t]
  \centering
  \includegraphics[width=\textwidth]{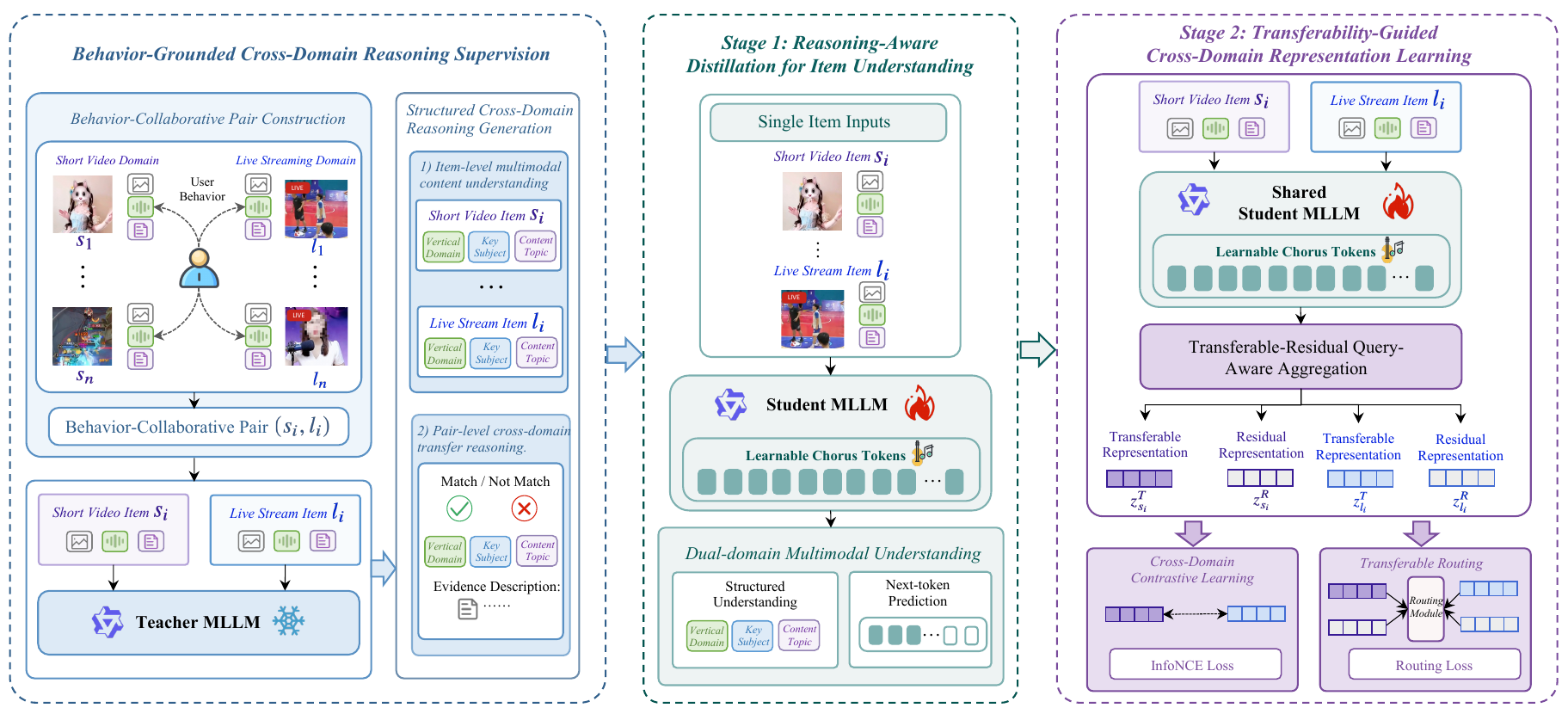}
  \caption{Overview of the proposed RGCD-Rep framework.}
  \label{fig:framework}
\end{figure*}

RGCD-Rep is a two-stage cross-domain representation learning framework,
as shown in Figure~\ref{fig:framework}.
It aims to introduce MLLM multimodal understanding and reasoning resource-efficiently while learning
offline-deployable cross-domain transferable item representations.
We first construct cross-domain item pairs from real behavioral logs,
and a frozen teacher MLLM generates structured transfer reasoning knowledge
as structured cross-domain reasoning supervision.
In Stage 1,
item-level content understanding knowledge is distilled into a lightweight student MLLM,
enabling multimodal understanding across both domains.
In Stage 2,
the student MLLM serves as a shared encoder and,
with transferable-residual query-aware aggregation,
learns cross-domain transferable and domain residual representations.
Together with cross-domain contrastive learning and transferable routing,
this stage models pair-level transfer relationships and mitigates negative transfer.

\subsection{Behavior-Grounded Cross-Domain Reasoning Supervision}
\subsubsection{Behavior-Collaborative Pair Construction}
\label{sec:pair_construction}
A key challenge in cross-domain transfer from short videos to live streaming is identifying item pairs
that reflect users' transferable interests.
We construct behavior-collaborative pairs from user interaction logs
by collecting each user's behaviors in both domains.
Intuitively,
if a user shows strong interest in a short video and a live stream within the same time window,
the two items likely share latent preference factors from the user's perspective.

Formally, we construct the behavior-collaborative pair dataset as
\begin{equation}
\mathcal{D}=\{(s_i,l_i,b_i)\}_{i=1}^{N_D},
\end{equation}
where $s_i$ denotes a short video,
$l_i$ denotes a live stream,
and $b_i$ denotes pair transferability strength.
The strength $b_i$ is computed from two signals.
The first is behavioral co-occurrence,
which measures how frequently $s_i$ and $l_i$ co-occur among the same user group within the same time window.
The second is behavioral intensity,
which measures user interaction strength with the two items.
For short videos,
we consider watch duration,
completion rate,
likes,
comments,
and shares.
For live streams,
we consider watch duration,
virtual-gift interactions,
follows,
likes,
and comments.

Finally,
we retain high-confidence behavioral pairs according to $b_i$,
which serve as training data for both Stage 1 and Stage 2.

\subsubsection{Structured Cross-Domain Reasoning Generation}
We use a frozen 35B MLLM as the teacher model to generate structured cross-domain transfer reasoning knowledge
for the behavior-collaborative pairs constructed in Section~\ref{sec:pair_construction}.
For each pair $(s_i, l_i)$,
the teacher MLLM takes the multimodal content of both items as input
and is prompted to analyze whether the user interests reflected in the short video
can be transferred to the live stream.

Instead of directly using free-form CoT text as supervision,
we convert the reasoning results of the teacher model into structured supervision signals.
Specifically,
the teacher model generates two types of reasoning knowledge.
The first type is \textit{item-level multimodal content understanding}.
For the short video $s_i$ and the live stream $l_i$,
the teacher model decomposes the content into three semantic dimensions:
\begin{equation}
\mathcal{K} = \{\text{vertical domain}, \text{key subject}, \text{content topic}\}.
\end{equation}
For each semantic dimension $k \in \mathcal{K}$,
the teacher model outputs a semantic label and a brief evidence description.
The second type is \textit{pair-level cross-domain transfer reasoning}.
For each semantic dimension $k$,
the teacher model determines whether the short video and the live stream match in this dimension:
\begin{equation}
m_i^k \in \{0,1\},
\end{equation}
where $m_i^k=1$ indicates that the two items match in semantic dimension $k$,
and $m_i^k=0$ indicates that they do not match.
The teacher model also generates a brief rationale for each matching decision.
Furthermore,
we aggregate the matching labels across the three dimensions
into a semantic transferability score from the perspective of teacher reasoning:
\begin{equation}
r_i = \frac{1}{|\mathcal{K}|} \sum_{k \in \mathcal{K}} m_i^k.
\end{equation}
This score measures the semantic transferability strength of pair $(s_i, l_i)$
from the reasoning perspective of the teacher model.

\subsection{Stage 1: Reasoning-Aware Distillation for Item Understanding}
\subsubsection{Pair-to-Item Reasoning Decomposition}
\label{sec:pair_to_item_decomposition}
Although the teacher MLLM reasons over short video and live stream pairs,
directly training the student model with pairwise inputs would reduce its deployability
in industrial recommendation systems. Therefore,
we decompose teacher-generated pair-level reasoning knowledge
into item-level distillation datasets $\mathcal{D}_{\text{item}}$.

For each pair $(s_i,l_i)$,
the item-level multimodal content understanding results generated by the teacher model
are decomposed into two independent supervision samples for the $s_i$ and the $l_i$,
respectively.
Each sample contains semantic labels and evidence descriptions
corresponding to the three semantic dimensions in $\mathcal{K}$.

Through this decomposition,
the student MLLM acquires multimodal content understanding capability from a dual-domain perspective.
Meanwhile,
it can independently encode any short video or live stream,
supporting large-scale offline embedding generation.

\subsubsection{Student MLLM Backbone with Learnable Chorus Tokens}
RGCD-Rep adopts a lightweight MLLM equipped with learnable chorus tokens as the student backbone,
following the design of CREM~\cite{LiuWangYangEtAl2026CREM}.
The core idea is to insert a compact set of learnable tokens into the multimodal prompt,
allowing the model to aggregate multimodal semantics from visual,
audio,
and textual signals into a small number of representation tokens while preserving its original generation capability.
Compared with directly adopting a single pooled token or the $\mathrm{EOS}$ token as the item representation,
chorus tokens provide a richer and more flexible representation interface
for multimodal understanding and downstream embedding-based tasks.

Formally,
given an item $x$,
its multimodal input contains video frames,
OCR/ASR text,
the title,
and other metadata.
We insert $N_C$ learnable chorus tokens into the input sequence:
\begin{equation}
C = [c_1, c_2, \cdots, c_{N_C}],
\end{equation}
where $c_i$ denotes the $i$-th chorus token.
The student MLLM jointly encodes the multimodal input and chorus tokens,
and outputs the hidden states of these chorus tokens from the last layer:
\begin{equation}
H_x = [h_x^1, h_x^2, \cdots, h_x^{N_C}] \in \mathbb{R}^{{N_C} \times d},
\end{equation}
where $h_x^i$ denotes the hidden state corresponding to the $i$-th chorus token,
and $d$ denotes the hidden dimension.
Following the default setting of CREM,
we set $N_C$ to $16$ in this paper.

\subsubsection{Student MLLM Distillation Objective}
The student model is a trainable lightweight 2B MLLM initialized from a general-purpose multimodal backbone.
The item-level multimodal content understanding results generated by the teacher MLLM are then converted into fine-tuning samples for the student model.

For each item $x$,
where $x$ can be either a short video or a live stream,
the input prompt consists of the multimodal content of the item
and a task instruction that asks the model to extract the vertical category,
key subject,
and content theme of the item.
The target answer is the item-level multimodal content understanding result generated by the teacher MLLM,
as described in Section~\ref{sec:pair_to_item_decomposition}.
Formally,
each sample is represented as $(q_x,y_x)$,
where $q_x$ denotes the multimodal instruction input for item $x$,
and $y_x=\{y_1,y_2,\cdots,y_{A_x}\}$ denotes the target answer sequence generated by the teacher model.

The student MLLM is optimized with the standard next-token prediction objective.
Given the input prompt $q_x$,
the chorus tokens $C$,
and the previously generated target tokens $y_{<t}$,
the student model predicts the next target token $y_t$:
\begin{equation}
\mathcal{L}_{\mathrm{NTP}}
=
-\sum_{x\in\mathcal{D}_{\mathrm{item}}}
\sum_{t=1}^{A_x}
\log P_{\theta}\left(y_t \mid q_x, C,y_{<t}\right),
\end{equation}
where $\theta$ denotes the parameters of the student MLLM,
$A_x$ denotes the length of the target answer sequence for item $x$,
and $\mathcal{D}_{\mathrm{item}}$ denotes the item-level distillation dataset.
After Stage 1 training,
the student MLLM acquires structured multimodal understanding capabilities for both short videos
and live streams without relying on the teacher model.

\subsection{Stage 2: Transferability-Guided Cross-Domain Representation Learning}
\subsubsection{Weight-Sharing MLLM Encoder}
In Stage 2,
the student MLLM distilled in Stage 1 serves as the encoding backbone,
and is shared by the short video and live streaming domains.
This design has two benefits.
First,
it maps short videos and live streams into a unified semantic space.
Second,
it transfers the structured multimodal understanding capability learned in Stage 1
to the representation learning process in Stage 2.
Specifically,
\begin{equation}
H_s=f_{\theta}(s_i), \quad H_l=f_{\theta}(l_i),
\end{equation}
where $f_{\theta}(\cdot)$ denotes the weight-shared student MLLM encoder.
Rather than relying on a single pooled representation,
we take the hidden states of the chorus tokens in the last layer as item representations.
Therefore,
for the item $x$,
the encoder output is represented as:
\begin{equation}
H_x=[h_x^1,h_x^2,\cdots,h_x^{N_C}] \in \mathbb{R}^{N_C \times d},
\end{equation}
where $h_x^c$ denotes the hidden state of the $c$-th special chorus token,
and $d$ denotes the hidden dimension.

\subsubsection{Transferable-Residual Query-Aware Aggregation}
Although the weight-shared MLLM encoder maps items from both domains into a unified semantic space,
chorus-token representations may still entangle transferable and domain-specific factors.
Therefore,
we further extract domain-transferable and domain residual representations from the chorus tokens.

Given the chorus-token representation $H_x$ of item $x$,
two independent learnable query vectors perform weighted aggregation over the token sequence,
yielding two semantic views:
the transferable view and the residual view.
For the transferable view,
a learnable query vector $q^T\in\mathbb{R}^{d}$ computes token attention weights:
\begin{equation}
\alpha_x^T
=
\operatorname{softmax}\left(\frac{H_x q^T}{\sqrt{d}}\right)
\in \mathbb{R}^{N_C},
\end{equation}
and the transferable representation is obtained by weighted pooling:
\begin{equation}
z_x^T
=
\operatorname{Norm}\left(
\sum_{c=1}^{N_C}\alpha_{x,c}^T h_x^c
\right).
\end{equation}
Similarly,
for the residual view,
another learnable query vector $q^R\in\mathbb{R}^{d}$ computes token weights:
\begin{equation}
\alpha_x^R
=
\operatorname{softmax}\left(\frac{H_x q^R}{\sqrt{d}}\right)
\in \mathbb{R}^{N_C},
\end{equation}
\begin{equation}
z_x^R
=
\operatorname{Norm}\left(
\sum_{c=1}^{N_C}\alpha_{x,c}^R h_x^c
\right).
\end{equation}

The module outputs the transferable representation $z_x^T$ and residual representation $z_x^R$ for each item.
Its parameters are instantiated separately for the short video and live streaming domains,
while shared among items within each domain.

\subsubsection{Cross-Domain Contrastive Learning}
Cross-domain contrastive learning optimizes the transferable embeddings.
Given a mini-batch $\mathcal{B}=\{(s_i,l_i)\}_{i=1}^{B}$,
the transferable embeddings $(z_{s_i}^{T},z_{l_i}^{T})$ of each behavior-collaborative pair
are treated as a positive pair.
For each item,
the unmatched items from the opposite domain in the same mini-batch serve as in-batch negative samples.
The contrastive loss from the short video domain to the live streaming domain is defined as
\begin{equation}
\mathcal{L}_{s \rightarrow l}
=
-\frac{1}{|\mathcal{B}|}
\sum_{i=1}^{|\mathcal{B}|}
\log
\frac{
\exp\left(\operatorname{sim}(z_{s_i}^{T},z_{l_i}^{T})/\tau\right)
}{
\sum_{j=1}^{|\mathcal{B}|}
\exp\left(\operatorname{sim}(z_{s_i}^{T},z_{l_j}^{T})/\tau\right)
}.
\end{equation}
Similarly,
the contrastive loss from the live streaming domain to the short video domain is defined as
\begin{equation}
\mathcal{L}_{l \rightarrow s}
=
-\frac{1}{|\mathcal{B}|}
\sum_{i=1}^{|\mathcal{B}|}
\log
\frac{
\exp\left(\operatorname{sim}(z_{l_i}^{T},z_{s_i}^{T})/\tau\right)
}{
\sum_{j=1}^{|\mathcal{B}|}
\exp\left(\operatorname{sim}(z_{l_i}^{T},z_{s_j}^{T})/\tau\right)
}.
\end{equation}
The final cross-domain contrastive loss is
\begin{equation}
\mathcal{L}_{\mathrm{con}}
=
\frac{1}{2}\left(
\mathcal{L}_{s \rightarrow l}
+
\mathcal{L}_{l \rightarrow s}
\right).
\end{equation}
Here,
$\operatorname{sim}(\cdot,\cdot)$ denotes cosine similarity,
and $\tau$ denotes the temperature coefficient.

\subsubsection{Transferable Routing}
Behavioral co-occurrence alone does not ensure true cross-domain transferability.
For low-transferability pairs,
pulling their transferable embeddings together may cause negative transfer.
To address this,
we introduce a transferable routing objective that determines whether a pair should be mainly explained
by the transferable or residual space.
For a pair $(s_i,l_i)$,
we compute its matching scores in the two spaces:
\begin{equation}
s_T(s_i,l_i)=\left(z_{s_i}^{T}\right)^\top z_{l_i}^{T},
\end{equation}
\begin{equation}
s_R(s_i,l_i)=\left(z_{s_i}^{R}\right)^\top z_{l_i}^{R}.
\end{equation}
The probability that this pair should be explained by the transferable space is
\begin{equation}
P_T(s_i,l_i)
=
\sigma\left(\gamma\left(s_T(s_i,l_i)-s_R(s_i,l_i)\right)\right),
\end{equation}
where $\sigma(\cdot)$ denotes the sigmoid function,
and $\gamma$ is the routing-logit scaling coefficient.

The routing label \(\pi_i\) is constructed from both the behavioral
transfer strength \(b_i\) and the semantic transferability score \(r_i\).
Specifically, we first compute a joint transferability score:
\begin{equation}
q_i = b_i + r_i .
\end{equation}

We then rank all training pairs according to \(q_i\). Let \(\mathcal{T}^{+}\)
and \(\mathcal{T}^{-}\) denote the top \(p\%\) and bottom \(p\%\) training
pairs ranked by \(q_i\), respectively. Pairs in \(\mathcal{T}^{+}\) are
treated as high-purity transferable pairs, while pairs in \(\mathcal{T}^{-}\)
are treated as low-purity pairs:
\begin{equation}
\pi_i =
\begin{cases}
1, & (s_i,l_i) \in \mathcal{T}^{+}, \\
0, & (s_i,l_i) \in \mathcal{T}^{-}.
\end{cases}
\end{equation}
The remaining pairs are regarded as uncertain samples and excluded from
the routing loss.
The routing loss is defined as
\begin{equation}
\begin{aligned}
\mathcal{L}_{\mathrm{route}}
&=
-\frac{1}{|\mathcal{B}_{\mathrm{route}}|}
\sum_{i\in\mathcal{B}_{\mathrm{route}}}
\left[
\pi_i \log P_T(s_i,l_i) \right. \\
&\left.\quad +
(1-\pi_i)\log\left(1-P_T(s_i,l_i)\right)
\right],
\end{aligned}
\end{equation}
where $\mathcal{B}_{\mathrm{route}}$ denotes high-purity and low-purity pairs
in the mini-batch participating in routing training.
For high-purity pairs,
the routing loss encourages $s_T(s_i,l_i)>s_R(s_i,l_i)$,
so the pair is mainly represented by transferable embeddings.
For low-purity pairs,
it encourages $s_R(s_i,l_i)>s_T(s_i,l_i)$,
allowing residual embeddings to absorb domain-specific factors or behavioral noise.

\subsubsection{Overall Training Objective}
The overall training objective of Stage 2 combines the cross-domain contrastive learning loss
and the transferable routing loss:
\begin{equation}
\mathcal{L}_{\mathrm{stage2}}
=
\mathcal{L}_{\mathrm{con}}
+
\lambda_{\mathrm{route}}\mathcal{L}_{\mathrm{route}},
\end{equation}
where $\lambda_{\mathrm{route}}$ controls the weight of the routing loss.
After training,
$z^T$ serves as the final cross-domain transferable item representation
and is passed to downstream recommendation tasks.

\section{Experiments}
In this section, we conduct extensive offline experiments and online A/B
tests on Kuaishou live streaming services, including Kuaishou and
Kuaishou Lite, to answer the following research questions:
\begin{list}{\labelitemi}{%
    \setlength{\leftmargin}{1.5em}%
    \setlength{\labelwidth}{1em}%
    \setlength{\labelsep}{0.5em}%
    \setlength{\itemsep}{0pt}%
    \setlength{\parsep}{0pt}%
}
    \item RQ1: How does the proposed RGCD-Rep framework perform compared
    with existing methods in the offline retrieval task?
    \item RQ2: How do the key modules of RGCD-Rep contribute?
    \item RQ3: Can RGCD-Rep improve long-tail recommendation?
    \item RQ4: Can RGCD-Rep benefit sparse-behavior users?
    \item RQ5: How does RGCD-Rep perform in online A/B tests?
\end{list}

\subsection{Experimental Settings}
\subsubsection{Kuaishou Dataset}
We conduct offline experiments on the Kuaishou dataset. Specifically, based
on real user behavior logs collected over one consecutive week
on the Kuaishou live streaming platform in 2026, we randomly sample
$109{,}523$ users who access both short video and live streaming services,
and record their historical positive interactions in both
domains. The dataset
consists of a training set and a test set.
The training set contains users' real positive interaction logs from the first six days, covering short videos and live streams observed during this training period.
The test set is constructed from the seventh day, where each user's last positive live streaming interaction on that day is used as the prediction target.
Each item contains multimodal information, including
video frames, OCR/ASR transcripts, title text, and other metadata. The
statistics of the datasets are reported in Table~\ref{tab:kuaishou_dataset}.
\begin{table}[t]
    \centering
    \caption{Statistics of the Kuaishou dataset.}
    \label{tab:kuaishou_dataset}
    \small
    \setlength{\tabcolsep}{4pt}
    \begin{tabular}{lrrrr}
        \toprule
        Domain & Users & Items & Interactions & Avg. Len. \\
        \midrule
        Live streaming & 109,523 & 98,190 & 2,275,888 & 20.78 \\
        Short video & 109,523 & 3,787,061 & 24,635,008 & 224.93 \\
        \bottomrule
    \end{tabular}
\end{table}

\subsubsection{Baselines}
To validate the effectiveness of RGCD-Rep, we compare it with three groups of baselines.
\begin{list}{\labelitemi}{%
    \setlength{\leftmargin}{1.5em}%
    \setlength{\labelwidth}{1em}%
    \setlength{\labelsep}{0.5em}%
    \setlength{\itemsep}{0pt}%
    \setlength{\parsep}{0pt}%
}
    \item \textbf{Single-domain multimodal recommendation methods}
    model user preferences or item representations within one domain,
    including FREEDOM~\cite{zhou2023freedom}, SMORE~\cite{ong2025smore},
    and AlphaRec~\cite{sheng2025language}.
    \item \textbf{Cross-domain unimodal recommendation methods} leverage
    information from different domains to alleviate data sparsity and
    cold-start issues in the target domain. However, these baselines
    mainly rely on unimodal representations as
    the cross-domain bridge, including UniSRec~\cite{hou2022unisrec}
    and VQ-Rec~\cite{hou2023vqrec}.
    \item \textbf{Cross-domain multimodal recommendation methods} learn
    transferable representations from item content such as text and visual
    information. However, existing methods directly adopt pre-extracted
    modality-specific representations as item features. These
    features are static and shallow, lack deep semantic reasoning over
    multimodal content. This category includes
    MISSRec~\cite{wang2023missrec}, PMMRec~\cite{li2024multimodality},
    and UniEmbedding~\cite{DaiEtAl2024UniEmbedding}.
\end{list}

\subsubsection{Implementation Details}
We implement RGCD-Rep in PyTorch on eight NVIDIA H800 (140GB) GPUs. The
teacher model is Qwen3.6-35B-A3B~\cite{qwen36_35b_a3b}, and the student model is
Qwen2-VL-2B-Instruct~\cite{Qwen2VL}. The number of fine-tuning samples is $60{,}000$ in
Stage 1 and $250{,}000$ in Stage 2. During fine-tuning, the LoRA rank
is set to $32$, and Adam serves as the optimizer. In Stage 1, the batch
size is $8$, and the learning rate is $1\times 10^{-4}$. In Stage 2, the
batch size is $64$, and the learning rate is $1\times 10^{-4}$. For the
loss function, the temperature coefficient is set to $\tau=0.05$, the
routing loss weight is set to $\lambda_{\mathrm{route}}=0.05$, the routing
logit scaling coefficient is set to $\gamma=10$. For routing-label construction, we set
\(p=20\). The final transferable embedding dimension is
$1536$. For all baselines, we carefully tune their hyperparameters to
obtain the best performance.

\subsubsection{Evaluation Metrics}
Following previous studies \cite{DaiEtAl2024UniEmbedding, YangYL2023MOTKD},
we evaluate RGCD-Rep and the baseline methods on the retrieval task, with
$\mathrm{HR}@K$ and $\mathrm{NDCG}@K$ as evaluation metrics, where $K=10$.

\subsection{Offline Performance on Matching Task (RQ1)}
\begin{table*}[t]
    \centering
    \caption{Offline performance on the matching task and ablation study.
    The best results are highlighted in bold, and the strongest baseline
    results are underlined.}
    \label{tab:offline_matching}
    \scriptsize
    \setlength{\tabcolsep}{2pt}
    \begin{tabular}{lcccccccccccc}
        \toprule
        Metric & FREEDOM & SMORE & AlphaRec & UniSRec & VQ-Rec
        & MISSRec & PMMRec & UniEmbedding
        & \cellcolor{RGCDblue}\textbf{RGCD-Rep} & w/o Stage 1
        & w/o Stage 2 & w/o Stage 2 MLLM \\
        \midrule
        HR@10
        & 0.0040 & 0.0089 & 0.0117 & 0.0134 & 0.0087 & 0.0147 & 0.0089
        & \underline{0.0216} & \cellcolor{RGCDblue}\textbf{0.0254}
        & 0.0249 & 0.0101 & 0.0189 \\
        NDCG@10
        & 0.0018 & 0.0032 & 0.0050 & 0.0059 & 0.0041 & 0.0066 & 0.0040
        & \underline{0.0097} & \cellcolor{RGCDblue}\textbf{0.0127}
        & 0.0122 & 0.0046 & 0.0088 \\
        \bottomrule
    \end{tabular}
\end{table*}

Table~\ref{tab:offline_matching} compares RGCD-Rep with all baselines and
ablation variants on the Kuaishou Dataset. For the matching task, we
mean-pool the transferable representations of a user's interacted short
videos as the user interest representation, and retrieve live streams by
maximum inner product search over candidate transferable representations.
The results yield the following observations.

First, single-domain multimodal methods perform relatively weakly.
Although FREEDOM~\cite{zhou2023freedom}, SMORE~\cite{ong2025smore}, and
AlphaRec~\cite{sheng2025language} exploit multimodal item content, they
mainly target single-domain recommendation and lack explicit modeling of
interest transfer from short videos to live streaming. Thus, they struggle
to alleviate sparsity in live streaming recommendation.

Second, cross-domain multimodal methods achieve stronger overall
performance. MISSRec~\cite{wang2023missrec},
PMMRec~\cite{li2024multimodality}, and
UniEmbedding~\cite{DaiEtAl2024UniEmbedding} jointly exploit multimodal
item content and cross-domain behavioral collaboration signals, learning
more effective matching representations. Among all baselines,
UniEmbedding~\cite{DaiEtAl2024UniEmbedding} performs best, indicating
that combining multimodal content with cross-domain behavior signals helps
capture shared interest patterns between short videos and live streams.

Finally, RGCD-Rep achieves the best performance on all metrics and
significantly outperforms all baselines, with improvements of
$17.59\%$ and $30.93\%$ over the strongest baseline, respectively. This
demonstrates that RGCD-Rep learns cross-domain transferable item
representations better suited to the live streaming recommendation.
\subsection{Ablation Study (RQ2)}
To validate each component in RGCD-Rep, we conduct
ablation experiments on the matching task, with the results
reported in Table~\ref{tab:offline_matching}. First, we remove the Stage 1
training objective, denoted as w/o Stage 1. This variant causes a slight
performance drop, indicating that reasoning-aware distillation provides a
beneficial semantic initialization for CDR. Next, we
remove the Stage 2 training objective, denoted as w/o Stage 2. The model
performance decreases substantially, showing that the item-level
multimodal content understanding learned through Stage 1 distillation
alone cannot fully support recommendation from short videos to live streams.
In contrast, pair-wise cross-domain representation learning explicitly
aligns transferable interest relations between the two domains based on
behavior-collaborative samples, which serves as a key source of
performance improvement. Finally, we freeze the MLLM backbone in Stage 2
and only train the transferable-residual query-aware aggregation module,
denoted as w/o Stage 2 MLLM. Its performance is clearly lower than that
of the full model. This result suggests that fixed multimodal representations remain relatively static for CDR and fail to
sufficiently model deep semantic associations and transferable relations
between short videos and live streams.

\subsection{Performance on Long-tail Items (RQ3)}
\begin{figure}[t]
    \centering
    \includegraphics[width=\columnwidth]{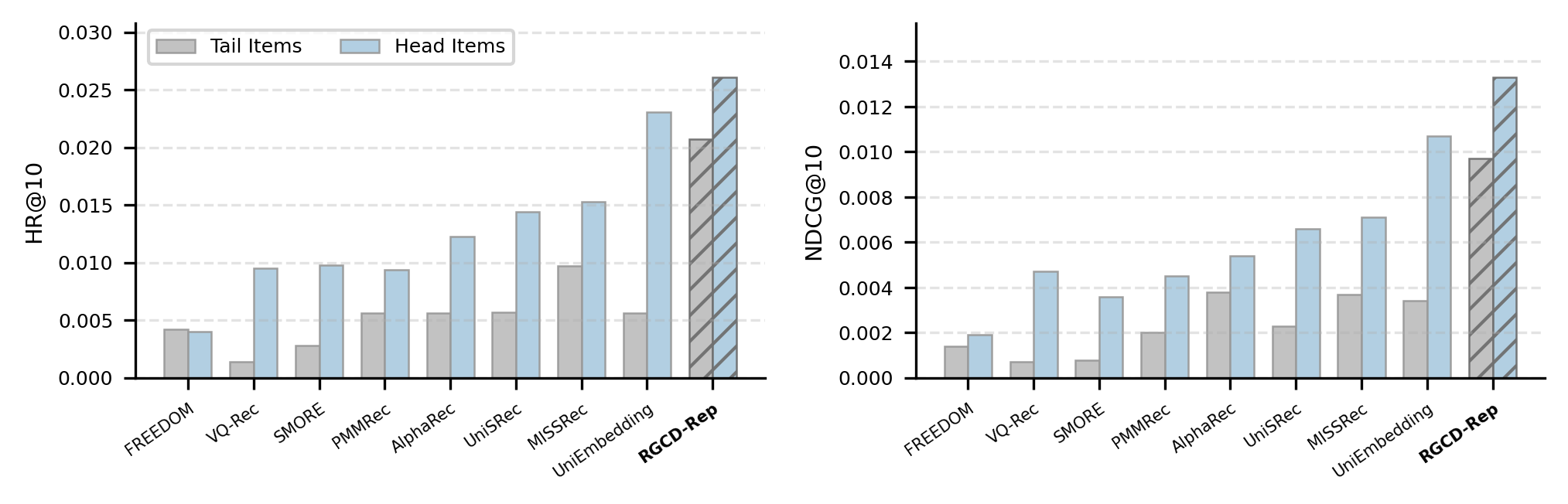}
    \caption{Performance comparison on tail and head items.}
    \label{fig:long_tail}
\end{figure}

We rank items by their occurrence frequency in the training set and label
the top $50\%$ as head items and the remaining items as tail items.
Interactions with tail items account for only $10.90\%$ of all
interactions, indicating a clear interaction sparsity issue in live
streaming recommendation. Figure~\ref{fig:long_tail} reports the
performance of different methods on head and tail items. All methods
perform worse on tail items than on head items, showing that long-tail
recommendation is more challenging under sparse interactions. RGCD-Rep
achieves the best performance on both head and tail items, with a more
pronounced advantage on tail items. This result suggests that RGCD-Rep can
maintain more reliable recommendation performance in sparse long-tail
scenarios by jointly modeling collaborative information and multimodal
information.

\subsection{Performance across User Groups (RQ4)}
\begin{figure}[t]
    \centering
    \includegraphics[width=\columnwidth]{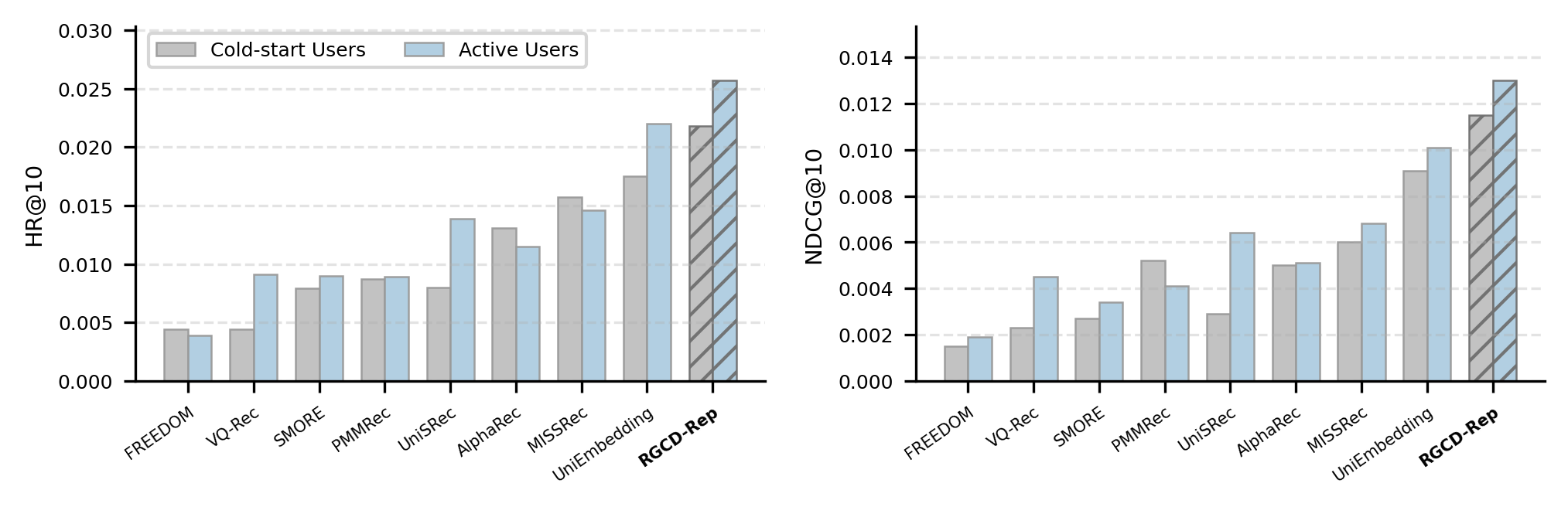}
    \caption{Performance comparison on cold-start and active users.}
    \label{fig:user_groups}
\end{figure}

To further evaluate RGCD-Rep's ability to alleviate behavioral
sparsity in the target live streaming domain, we divide test users into
cold-start users and active users based on their historical live streaming feedback.
Specifically, a user is regarded as cold-start if they have no positive live streaming feedback during the six days before the prediction day and generate positive feedback only on the prediction day; otherwise, the user is regarded as active.
Cold-start users account for $14.65\%$ of the test users.
Figure~\ref{fig:user_groups} reports the results of
different methods on the two user groups.
RGCD-Rep outperforms all baselines for both
cold-start users and active users, indicating that it improves
recommendation performance not only for active users but also for users with sparse live streaming behaviors.
More importantly, the improvement of RGCD-Rep over the strongest baseline is
more pronounced for cold-start users than for active users, suggesting
that its performance gain does not merely come from modeling existing
live streaming behaviors. By effectively transferring user interests from
the short video domain, RGCD-Rep supplements user preference signals when
live streaming behaviors are sparse, thereby alleviating the user-side
cold-start problem in the live streaming domain.

\subsection{Online A/B Tests (RQ5)}
\begin{table}[t]
    \centering
    \caption{Online A/B test performance in live streaming services of Kuaishou and Kuaishou Lite.}
    \label{tab:online_ab}
    \scriptsize
    \setlength{\tabcolsep}{2pt}
    \begin{tabular}{lcccc}
        \toprule
        Application & Exposure & Click & \shortstack{Effective Room-entry} & Follow \\
        \midrule
        \textbf{Kuaishou} & +0.37\% & +0.34\% & +0.28\% & +0.83\% \\
        \textbf{Kuaishou Lite} & +0.38\% & +0.13\% & +0.22\% & +0.93\% \\
        \bottomrule
    \end{tabular}
\end{table}
We deploy RGCD-Rep to the live streaming recommendation scenarios of
Kuaishou and Kuaishou Lite, serving hundreds of millions of users at the
retrieval stage, and conduct a one-week online A/B test. For deployment,
we rebuild the training data from recent production logs and select about
one million high-quality pairs to train the student MLLM. After training,
the student MLLM offline generates transferable representations for all
short videos and live streams, which are stored in the item feature
system. During online retrieval, we build a cross-domain nearest-neighbor
index from these offline embeddings and retrieve transferable live streams
candidates based on users' short video interests as a new retrieval
channel.

As shown in Table~\ref{tab:online_ab}, RGCD-Rep brings significant gains
on both applications. Follows improve most notably, reaching $+0.83\%$
and $+0.93\%$, respectively, while core metrics such as effective
room-entry rate and clicks also increase steadily. These
results show that introducing MLLM-based cross-domain reasoning into
transferable representation learning can accurately capture transfer
relations from short video interests to live streamers, especially
improving users' following of new streamers and enabling more efficient
live streaming content distribution.

\section{Conclusion}
This paper proposes RGCD-Rep, a reasoning-guided cross-domain representation
learning framework for short video to live streaming recommendation that
incorporates MLLM multimodal understanding and cross-domain reasoning into
transferable item representation learning. The transferable item representations can be generated offline and integrated into downstream recommendation systems, meeting industrial low-cost deployment requirements while leveraging MLLM reasoning.
Extensive offline experiments and online A/B tests show that RGCD-Rep brings consistent
improvements in live streaming matching and sparse scenarios such as long-tail
and cold-start recommendation. RGCD-Rep is deployed at Kuaishou, serving over
$400$ million users in real-world live streaming services.



\bibliographystyle{ACM-Reference-Format}
\bibliography{sample-base}

@String{Computing = "Computing" }

@inproceedings{guo2026room,
  author    = {Ke Guo and Changle Qu and Xiao Zhang and Liqin Zhao and Shijun Wang and Yanan Niu and Jun Xu},
  title     = {Room Matters: Dynamic Room‐level Collaboration Information Modeling for Live Streaming Recommendation},
  booktitle = {Proceedings of The Web Conference 2026},
  year      = {2026},
  pages     = {6045--6056},
  publisher = {ACM},
  doi       = {10.1145/3726302.3729914},
  url       = {https://doi.org/10.1145/3774904.3792241}
}

@inproceedings{qu2025bridging,
  author    = {Changle Qu and Liqin Zhao and Yanan Niu and Xiao Zhang and Jun Xu},
  title     = {Bridging Short Videos and Streamers with Multi-Graph Contrastive Learning for Live Streaming Recommendation},
  booktitle = {Proceedings of the 48th International ACM SIGIR Conference on Research and Development in Information Retrieval (SIGIR '25)},
  year      = {2025},
  pages     = {2059--2069},
  publisher = {ACM},
  doi       = {10.1145/3726302.3729914},
  url       = {https://doi.org/10.1145/3726302.3729914}
}

@inproceedings{cao2026foresight,
  author    = {Jiangxia Cao and Ruochen Yang and Xiang Chen and Changxin Lao and Yueyang Liu and Yusheng Huang and Yuanhao Tian and Xiangyu Wu and Shuang Yang and Zhaojie Liu and Guorui Zhou},
  title     = {Foresight Prediction Enhanced Live-Streaming Recommendation},
  booktitle = {Proceedings of the Nineteenth ACM International Conference on Web Search and Data Mining (WSDM '26)},
  year      = {2026},
  publisher = {ACM},
  doi       = {10.1145/3773966.3779372},
  url       = {https://doi.org/10.1145/3773966.3779372}
}

@inproceedings{li2024aiming,
  author    = {Hanyu Li and Weizhi Ma and Peijie Sun and Jiayu Li and Cunxiang Yin and Yancheng He and Guoqiang Xu and Min Zhang and Shaoping Ma},
  title     = {Aiming at the Target: Filter Collaborative Information for Cross-Domain Recommendation},
  booktitle = {Proceedings of the 47th International ACM SIGIR Conference on Research and Development in Information Retrieval (SIGIR '24)},
  year      = {2024},
  pages     = {2081--2090},
  publisher = {ACM},
  doi       = {10.1145/3626772.3657713},
  url       = {https://doi.org/10.1145/3626772.3657713}
}

@inproceedings{li2025disco,
  title={DisCo: graph-based disentangled contrastive learning for cold-start cross-domain recommendation},
  author={Li, Hourun and Wang, Yifan and Xiao, Zhiping and Yang, Jia and Zhou, Changling and Zhang, Ming and Ju, Wei},
  booktitle={Proceedings of the AAAI Conference on Artificial Intelligence},
  volume={39},
  number={11},
  pages={12049--12057},
  year={2025}
}

@inproceedings{zhao2023crossdomain,
  author    = {Chuang Zhao and Hongke Zhao and Ming He and Jian Zhang and Jianping Fan},
  title     = {Cross‑domain recommendation via user interest alignment},
  booktitle = {Proceedings of The Web Conference 2023 (WWW '23)},
  year      = {2023},
  pages     = {887--896},
  publisher = {ACM},
  doi       = {10.1145/3543507.3583263},
  url       = {https://doi.org/10.1145/3543507.3583263}
}

@inproceedings{DaiEtAl2024UniEmbedding,
  title        = {UniEmbedding: Learning Universal Multi-Modal Multi-Domain Item Embeddings via User-View Contrastive Learning},
  author       = {Dai, Boqi and Du, Zhaocheng and Zhu, Jieming and Xu, Jintao and Zou, Deqing and Dai, Quanyu and Dong, Zhenhua and Zhang, Rui and Zheng, Hai-Tao},
  booktitle    = {Proceedings of the 33rd ACM International Conference on Information and Knowledge Management (CIKM '24)},
  year         = {2024},
  pages        = {4446--4453},
  publisher    = {ACM},
  doi          = {10.1145/3627673.3680098},
  url          = {https://doi.org/10.1145/3627673.3680098}
}

@inproceedings{HuZY18CoNet,
  title        = {CoNet: Collaborative Cross Networks for Cross-Domain Recommendation},
  author       = {Guangneng Hu and Yu Zhang and Qiang Yang},
  booktitle    = {Proceedings of the 27th ACM International Conference on Information and Knowledge Management (CIKM 2018)},
  year         = {2018},
  pages        = {667--676},
  publisher    = {ACM},
  doi          = {10.1145/3269206.3271684},
  url          = {https://doi.org/10.1145/3269206.3271684},
  isbn         = {978-1-4503-6014-2}
}

@inproceedings{LiuLLP20BiTGCF,
  title        = {Cross Domain Recommendation via Bi-directional Transfer Graph Collaborative Filtering Networks},
  author       = {Liu, Meng and Li, Jianjun and Li, Guohui and Pan, Peng},
  booktitle    = {Proceedings of the 29th ACM International Conference on Information and Knowledge Management (CIKM '20)},
  year         = {2020},
  pages        = {885--894},
  publisher    = {ACM},
  doi          = {10.1145/3340531.3412012},
  url          = {https://doi.org/10.1145/3340531.3412012}
}

@inproceedings{ManShenJinCheng2017CrossDomain,
  title        = {Cross-Domain Recommendation: An Embedding and Mapping Approach},
  author       = {Man, Tong and Shen, Huawei and Jin, Xiaolong and Cheng, Xueqi},
  booktitle    = {Proceedings of the Twenty-Sixth International Joint Conference on Artificial Intelligence (IJCAI-17)},
  year         = {2017},
  pages        = {2464--2470},
  publisher    = {IJCAI Organization},
  doi          = {10.24963/ijcai.2017/343},
  url          = {https://doi.org/10.24963/ijcai.2017/343}
}

@inproceedings{XieLiuWangLiuZhangLin2022CCDR,
  title        = {Contrastive Cross-Domain Recommendation in Matching},
  author       = {Xie, Ruobing and Liu, Qi and Wang, Liangdong and Liu, Shukai and Zhang, Bo and Lin, Leyu},
  booktitle    = {Proceedings of the 28th ACM SIGKDD Conference on Knowledge Discovery and Data Mining (KDD '22)},
  year         = {2022},
  pages        = {4226--4236},
  publisher    = {ACM},
  doi          = {10.1145/3534678.3539125},
  url          = {https://doi.org/10.1145/3534678.3539125}
}

@inproceedings{CaoCongShengLiuWang2022C2DSR,
  title        = {Contrastive Cross-Domain Sequential Recommendation},
  author       = {Cao, Jiangxia and Cong, Xin and Sheng, Jiawei and Liu, Tingwen and Wang, Bin},
  booktitle    = {Proceedings of the 31st ACM International Conference on Information and Knowledge Management (CIKM '22)},
  year         = {2022},
  pages        = {138--147},
  publisher    = {ACM},
  doi          = {10.1145/3511808.3557262},
  url          = {https://doi.org/10.1145/3511808.3557262}
}

@inproceedings{HuXiaZhangFuWuHuanLiTangZhou2024EnhancingSR,
  title        = {Enhancing Sequential Recommendation via LLM-based Semantic Embedding Learning},
  author       = {Hu, Jun and Xia, Wenwen and Zhang, Xiaolu and Fu, Chilin and Wu, Weichang and Huan, Zhaoxin and Li, Ang and Tang, Zuoli and Zhou, Jun},
  booktitle    = {Proceedings of The Web Conference 2024 (WWW '24)},
  year         = {2024},
  pages        = {103--111},
  publisher    = {ACM},
  doi          = {10.1145/3589335.3648307},
  url          = {https://doi.org/10.1145/3589335.3648307}
}

@inproceedings{YangYL2023MOTKD,
  title        = {Multimodal Optimal Transport Knowledge Distillation for Cross-domain Recommendation},
  author       = {Yang, Wei and Yang, Jie and Liu, Yuan},
  booktitle    = {Proceedings of the 32nd ACM International Conference on Information and Knowledge Management (CIKM '23)},
  year         = {2023},
  pages        = {2959--2968},
  publisher    = {ACM},
  doi          = {10.1145/3583780.3614983},
  url          = {https://doi.org/10.1145/3583780.3614983}
}

@article{LiuWangYangEtAl2026CREM,
  title        = {CREM: Compression-Driven Representation Enhancement for Multimodal Retrieval and Comprehension},
  author       = {Liu, Lihao and Wang, Yan and Yang, Biao and Li, Da and Cao, Jiangxia and Luo, Yuxiao and Chen, Xiang and Wu, Xiangyu and Yuan, Wei and Yang, Fan and Ding, Guiguang and Gao, Tingting and Zhou, Guorui},
  journal      = {arXiv preprint arXiv:2602.19091},
  year         = {2026},
  url          = {https://arxiv.org/abs/2602.19091},
  note         = {arXiv:2602.19091}
}

@article{LiYangWenEtAl2025FARM,
  title        = {FARM: Frequency-Aware Model for Cross-Domain Live-Streaming Recommendation},
  author       = {Li, Xiaodong and Yang, Ruochen and Wen, Shuang and Wang, Shen and Liu, Yueyang and Wang, Guoquan and Hu, Weisong and Luo, Qiang and Sheng, Jiawei and Liu, Tingwen and Cao, Jiangxia and Yang, Shuang and Liu, Zhaojie},
  journal      = {CoRR},
  volume       = {abs/2502.09375},
  year         = {2025},
  doi          = {10.48550/arXiv.2502.09375},
  url          = {https://arxiv.org/abs/2502.09375},
  note         = {arXiv preprint}
}

@inproceedings{li2024multimodality,
  title        = {Multi-Modality is All You Need for Transferable Recommender Systems},
  author       = {Li, Youhua and Du, Hanwen and Ni, Yongxin and Zhao, Pengpeng and Guo, Qi and Yuan, Fajie and Zhou, Xiaofang},
  booktitle    = {2024 IEEE 40th International Conference on Data Engineering (ICDE)},
  year         = {2024},
  pages        = {5008--5021},
  publisher    = {IEEE},
  doi          = {10.1109/ICDE60146.2024.00380},
  url          = {https://doi.org/10.1109/ICDE60146.2024.00380}
}

@inproceedings{wang2023missrec,
  title        = {{MISSRec}: Pre-training and Transferring Multi-modal Interest-aware Sequence Representation for Recommendation},
  author       = {Wang, Jinpeng and Zeng, Ziyun and Wang, Yunxiao and Wang, Yuting and Lu, Xingyu and Li, Tianxiang and Yuan, Jun and Zhang, Rui and Zheng, Hai-Tao and Xia, Shu-Tao},
  booktitle    = {Proceedings of the 31st ACM International Conference on Multimedia},
  series       = {MM '23},
  year         = {2023},
  pages        = {6548--6557},
  publisher    = {Association for Computing Machinery},
  address      = {New York, NY, USA},
  doi          = {10.1145/3581783.3611967},
  url          = {https://doi.org/10.1145/3581783.3611967}
}

@inproceedings{hou2023vqrec,
  title        = {Learning Vector-Quantized Item Representation for Transferable Sequential Recommenders},
  author       = {Hou, Yupeng and He, Zhankui and McAuley, Julian and Zhao, Wayne Xin},
  booktitle    = {Proceedings of the ACM Web Conference 2023},
  series       = {WWW '23},
  year         = {2023},
  pages        = {1162--1171},
  publisher    = {Association for Computing Machinery},
  address      = {New York, NY, USA},
  doi          = {10.1145/3543507.3583434},
  url          = {https://doi.org/10.1145/3543507.3583434}
}

@inproceedings{hou2022unisrec,
  title        = {Towards Universal Sequence Representation Learning for Recommender Systems},
  author       = {Hou, Yupeng and Mu, Shanlei and Zhao, Wayne Xin and Li, Yaliang and Ding, Bolin and Wen, Ji-Rong},
  booktitle    = {Proceedings of the 28th ACM SIGKDD Conference on Knowledge Discovery and Data Mining},
  series       = {KDD '22},
  year         = {2022},
  pages        = {585--593},
  publisher    = {Association for Computing Machinery},
  address      = {New York, NY, USA},
  doi          = {10.1145/3534678.3539381},
  url          = {https://doi.org/10.1145/3534678.3539381}
}

@inproceedings{sheng2025language,
  title        = {Language Representations Can be What Recommenders Need: Findings and Potentials},
  author       = {Sheng, Leheng and Zhang, An and Zhang, Yi and Chen, Yuxin and Wang, Xiang and Chua, Tat-Seng},
  booktitle    = {The Thirteenth International Conference on Learning Representations},
  year         = {2025},
  url          = {https://openreview.net/forum?id=eIJfOIMN9z}
}

@inproceedings{ong2025smore,
  title        = {Spectrum-based Modality Representation Fusion Graph Convolutional Network for Multimodal Recommendation},
  author       = {Ong, Rongqing Kenneth and Khong, Andy W. H.},
  booktitle    = {Proceedings of the Eighteenth ACM International Conference on Web Search and Data Mining},
  series       = {WSDM '25},
  year         = {2025},
  pages        = {773--781},
  publisher    = {Association for Computing Machinery},
  address      = {New York, NY, USA},
  doi          = {10.1145/3701551.3703561},
  url          = {https://doi.org/10.1145/3701551.3703561}
}

@inproceedings{zhou2023freedom,
  title        = {A Tale of Two Graphs: Freezing and Denoising Graph Structures for Multimodal Recommendation},
  author       = {Zhou, Xin and Shen, Zhiqi},
  booktitle    = {Proceedings of the 31st ACM International Conference on Multimedia},
  series       = {MM '23},
  year         = {2023},
  pages        = {935--943},
  publisher    = {Association for Computing Machinery},
  address      = {New York, NY, USA},
  doi          = {10.1145/3581783.3611943},
  url          = {https://doi.org/10.1145/3581783.3611943}
}

@inproceedings{deng2024mmbee,
  title        = {{MMBee}: Live Streaming Gift-Sending Recommendations via Multi-Modal Fusion and Behaviour Expansion},
  author       = {Deng, Jiaxin and Wang, Shiyao and Wang, Yuchen and Qi, Jiansong and Zhao, Liqin and Zhou, Guorui and Meng, Gaofeng},
  booktitle    = {Proceedings of the 30th ACM SIGKDD Conference on Knowledge Discovery and Data Mining},
  series       = {KDD '24},
  year         = {2024},
  pages        = {4896--4905},
  publisher    = {Association for Computing Machinery},
  address      = {New York, NY, USA},
  doi          = {10.1145/3637528.3671511},
  url          = {https://doi.org/10.1145/3637528.3671511}
}

@inproceedings{liu2023llava,
  title     = {Visual Instruction Tuning},
  author    = {Liu, Haotian and Li, Chunyuan and Wu, Qingyang and Lee, Yong Jae},
  booktitle = {Advances in Neural Information Processing Systems},
  volume    = {36},
  year      = {2023}
}

@inproceedings{dai2023instructblip,
  title     = {{InstructBLIP}: Towards General-purpose Vision-Language Models with Instruction Tuning},
  author    = {Dai, Wenliang and Li, Junnan and Li, Dongxu and Tiong, Anthony Meng Huat and Zhao, Junqi and Wang, Weisheng and Li, Boyang and Fung, Pascale and Hoi, Steven C. H.},
  booktitle = {Advances in Neural Information Processing Systems},
  volume    = {36},
  year      = {2023}
}

@inproceedings{caffagni2024revolution,
  title     = {The Revolution of Multimodal Large Language Models: A Survey},
  author    = {Caffagni, Davide and Cocchi, Federico and Barsellotti, Luca and Moratelli, Nicholas and Sarto, Sara and Baraldi, Lorenzo and Baraldi, Lorenzo and Cornia, Marcella and Cucchiara, Rita},
  booktitle = {Findings of the Association for Computational Linguistics: ACL 2024},
  year      = {2024},
  doi       = {10.18653/v1/2024.findings-acl.807}
}

@inproceedings{ye2025harnessing,
  title        = {Harnessing Multimodal Large Language Models for Multimodal Sequential Recommendation},
  author       = {Ye, Yuyang and Zheng, Zhi and Shen, Yishan and Wang, Tianshu and Zhang, Hengruo and Zhu, Peijun and Yu, Runlong and Zhang, Kai and Xiong, Hui},
  booktitle    = {Proceedings of the AAAI Conference on Artificial Intelligence},
  year         = {2025},
  volume       = {39},
  number       = {12},
  pages        = {13069--13077},
  doi          = {10.1609/aaai.v39i12.33426},
  url          = {https://doi.org/10.1609/aaai.v39i12.33426}
}

@inproceedings{zhu2022ptupcdr,
  title     = {Personalized Transfer of User Preferences for Cross-domain Recommendation},
  author    = {Zhu, Yongchun and Ge, Kaikai and Zhuang, Fuzhen and Xie, Ruobing and Xi, Dongbo and Zhang, Xu and Lin, Leyu and He, Qing},
  booktitle = {Proceedings of the Fifteenth ACM International Conference on Web Search and Data Mining},
  pages     = {1507--1515},
  year      = {2022},
  doi       = {10.1145/3488560.3498388}
}

@inproceedings{park2023negative,
  title     = {Cracking the Code of Negative Transfer: A Cooperative Game Theoretic Approach for Cross-Domain Sequential Recommendation},
  author    = {Park, Chung and Kim, Taesan and Choi, Taekyoon and Hong, Junui and Yu, Yelim and Cho, Mincheol and Lee, Kyunam and Ryu, Sungil and Yoon, Hyungjun and Choi, Minsung and Choo, Jaegul},
  booktitle = {Proceedings of the 32nd ACM International Conference on Information and Knowledge Management},
  pages     = {2024--2033},
  year      = {2023},
  doi       = {10.1145/3583780.3614828}
}

@article{ma2024tricdr,
  title        = {Triple Sequence Learning for Cross-domain Recommendation},
  author       = {Ma, Haokai and Xie, Ruobing and Meng, Lei and Chen, Xin and Zhang, Xu and Lin, Leyu and Zhou, Jie},
  journal      = {ACM Transactions on Information Systems},
  volume       = {42},
  number       = {4},
  articleno    = {91},
  pages        = {91:1--91:29},
  year         = {2024},
  publisher    = {Association for Computing Machinery},
  doi          = {10.1145/3638351},
  url          = {https://doi.org/10.1145/3638351}
}

@inproceedings{chen2017acf,
  title        = {Attentive Collaborative Filtering: Multimedia Recommendation with Item- and Component-Level Attention},
  author       = {Chen, Jingyuan and Zhang, Hanwang and He, Xiangnan and Nie, Liqiang and Liu, Wei and Chua, Tat-Seng},
  booktitle    = {Proceedings of the 40th International ACM SIGIR Conference on Research and Development in Information Retrieval},
  series       = {SIGIR '17},
  year         = {2017},
  pages        = {335--344},
  publisher    = {Association for Computing Machinery},
  address      = {New York, NY, USA},
  doi          = {10.1145/3077136.3080797},
  url          = {https://doi.org/10.1145/3077136.3080797}
}

@inproceedings{wei2019mmgcn,
  title        = {{MMGCN}: Multi-modal Graph Convolution Network for Personalized Recommendation of Micro-video},
  author       = {Wei, Yinwei and Wang, Xiang and Nie, Liqiang and He, Xiangnan and Hong, Richang and Chua, Tat-Seng},
  booktitle    = {Proceedings of the 27th ACM International Conference on Multimedia},
  series       = {MM '19},
  year         = {2019},
  pages        = {1437--1445},
  publisher    = {Association for Computing Machinery},
  address      = {New York, NY, USA},
  doi          = {10.1145/3343031.3351034},
  url          = {https://doi.org/10.1145/3343031.3351034}
}

@inproceedings{wei2020grcn,
  title        = {{GRCN}: Graph-Refined Convolutional Network for Multimedia Recommendation with Implicit Feedback},
  author       = {Wei, Yinwei and Wang, Xiang and Nie, Liqiang and He, Xiangnan and Chua, Tat-Seng},
  booktitle    = {Proceedings of the 28th ACM International Conference on Multimedia},
  series       = {MM '20},
  year         = {2020},
  pages        = {3541--3549},
  publisher    = {Association for Computing Machinery},
  address      = {New York, NY, USA},
  doi          = {10.1145/3394171.3413556},
  url          = {https://doi.org/10.1145/3394171.3413556}
}

@inproceedings{chen2019vecf,
  title        = {Personalized Fashion Recommendation with Visual Explanations based on Multimodal Attention Network: Towards Visually Explainable Recommendation},
  author       = {Chen, Xu and Chen, Hanxiong and Xu, Hongteng and Zhang, Yongfeng and Cao, Yixin and Qin, Zheng and Zha, Hongyuan},
  booktitle    = {Proceedings of the 42nd International ACM SIGIR Conference on Research and Development in Information Retrieval},
  series       = {SIGIR '19},
  year         = {2019},
  pages        = {765--774},
  publisher    = {Association for Computing Machinery},
  address      = {New York, NY, USA},
  doi          = {10.1145/3331184.3331254},
  url          = {https://doi.org/10.1145/3331184.3331254}
}

@misc{qwen36_35b_a3b,
    title = {{Qwen3.6-35B-A3B}: Agentic Coding Power, Now Open to All},
    url = {https://qwen.ai/blog?id=qwen3.6-35b-a3b},
    author = {{Qwen Team}},
    month = {April},
    year = {2026}
}

@article{Qwen2VL,
  title={Qwen2-VL: Enhancing Vision-Language Model's Perception of the World at Any Resolution},
  author={Wang, Peng and Bai, Shuai and Tan, Sinan and Wang, Shijie and Fan, Zhihao and Bai, Jinze and Chen, Keqin and Liu, Xuejing and Wang, Jialin and Ge, Wenbin and Fan, Yang and Dang, Kai and Du, Mengfei and Ren, Xuancheng and Men, Rui and Liu, Dayiheng and Zhou, Chang and Zhou, Jingren and Lin, Junyang},
  journal={arXiv preprint arXiv:2409.12191},
  year={2024}
}

\end{document}